\documentclass[12pt]{article}
\usepackage{axodraw2}
\usepackage{latexsym, graphicx,cite,subcaption,enumitem}
\usepackage{amsmath}
\usepackage{amssymb}
\usepackage{amsthm}

\usepackage[top = 1. in,bottom = 1. in, left = 1 in, right=1 in]{geometry}

 \newcommand{\arXiv}[1]{\href{http://www.arXiv.org/abs/#1}{arXiv:#1}}
\usepackage[colorlinks=true, linkcolor=blue, bookmarks=true]{hyperref}

\makeatletter
\renewcommand\section{\@startsection {section}{1}{\z@}%
                  {-3.5ex \@plus -1ex \@minus -.2ex}
                  {2.3ex \@plus.2ex}%
                  {\normalfont\large\bfseries}}
\renewcommand\subsection{\@startsection{subsection}{2}{\z@}%
                   {-3.25ex\@plus -1ex \@minus -.2ex}%
                   {1.5ex \@plus .2ex}%
                   {\normalfont\bfseries}}
\makeatother

\newcommand{\beq}{\begin{equation}}
\newcommand{\eeq}{\end{equation}}
\newcommand{\ber}{\begin{array}}
\newcommand{\eer}{\end{array}}

\newcommand{\de}{\delta}

\newcommand{\ena}{\end{eqnarray}}
\newcommand{\beqa}{\begin{eqnarray}}
\newcommand{\eeqa}{\end{eqnarray}}
\newcommand{\bea}{\begin{eqnarray}}
\newcommand{\eea}{\end{eqnarray}}

\theoremstyle{remark}

\begin{document}
\begin{titlepage}
\phantom{.}\vspace{1cm}

\begin{center}
{\LARGE\bf \mbox{Melonic dominance and the largest}\vspace{3mm}\\
eigenvalue of a large random tensor}\\
\vskip 15mm
{\large Oleg Evnin$^{a,b}$}
\vskip 7mm
{\em $^a$ Department of Physics, Faculty of Science, Chulalongkorn University,
Bangkok, Thailand}
\vskip 3mm
{\em $^b$ Theoretische Natuurkunde, Vrije Universiteit Brussel and\\
The International Solvay Institutes, Brussels, Belgium}
\vskip 7mm
{\small\noindent {\tt oleg.evnin@gmail.com}}
\vskip 10mm
\end{center}
\vspace{3cm}
\begin{center}
{\bf ABSTRACT}\vspace{3mm}
\end{center}

We consider a Gaussian rotationally invariant ensemble of random real totally symmetric tensors with independent normally distributed entries, and estimate the largest eigenvalue of a typical tensor in this ensemble by examining the rate of growth of a random initial vector under successive applications of a nonlinear map defined by the random tensor. In the limit of a large number of dimensions, we observe that a simple form of melonic dominance holds, and the quantity we study is effectively determined by a single Feynman diagram arising from the Gaussian average over the tensor components. This computation suggests that the largest tensor eigenvalue in our ensemble in the limit of a large number of dimensions is proportional to the square root of the number of dimensions, as it is for random real symmetric matrices.

\vfill

\end{titlepage}


\section{Introduction}

The aim of this letter is to call attention to the topic of eigenvalues of random tensors, a subject that appears to have largely fallen through the cracks between a few active but disjoint communities.

Statistics of eigenvalues plays a crucial role in the standard theory of random matrices \cite{mehta}. Indeed, the spectral decomposition of matrices allows for a straightforward extraction of the distribution of eigenvalues for the standard Gaussian ensembles of random matrices. Universal statistical properties emerge in these distributions when one deals with matrices of size $N\times N$ with a large $N$.

Random tensors \cite{melon3} have emerged as a generalization of random matrices, largely motivated by the needs of high-energy physics. Expectation values of various quantities in ensembles of Gaussian random tensors are represented graphically as Feynman diagrams. It turns out that for a tensor in a large number of dimensions $N$, a special class of diagrams called `melonic' dominates, which creates effective methodology for treating such statistical ensembles. Note, however, that, unlike for matrices, the eigenvalues of tensors are not evoked for such analysis.

The eigenvalues of tensors, without randomness, have received a significant amount of attention starting with \cite{Qi, Lim}, largely from the perspective of optimization problems, for reviews see \cite{Qirev,eigbook}.
The subject falls into the domain of nonlinear algebra \cite{DM}. There is no unique generalization of matrix eigenvalues to tensors, and a few relevant definitions have been proposed. The properties of eigenvalues and eigenvectors of tensors are much more intricate than for matrices, and relatively poorly understood. The number of eigenvalues is typically exponentially large in the number of dimensions \cite{DM,numbereig}. Some techniques for constructing the characteristic polynomials have been presented in \cite{MSh}. There is no obvious relation between the eigenvectors of tensors and decompositions analogous to the spectral decomposition of matrices, such as the CP-decomposition \cite{rank}, or its symmetric version \cite{symmrank}. (See \cite{robeva}, however, for a special case where such a relation can be established.)

What about eigenvalues of random tensors? In the absense of a reliable analog of spectral decomposition for matrices, and given the typically huge number of tensor eigenvalues, it is unrealistic to develop a parametrization of tensors based on their eigenvalues, as done for matrices, and hence effectively extract the joint distribution of the eigenvalues from the distributions of the tensor components. The strongest tool at our disposal\footnote{It is appropriate to mention a few past works dealing with related problems from a different perspective. Thus, \cite{ABC} analyzes the critical points of the spherical $p$-spin spin-glass model Hamiltonian that have many algebraic similarities to tensor eigenvalues, while a Gaussian random tensor appears in the definition of this Hamiltonian. In \cite{cooper}, the largest H-eigenvalue (rather than the E-eigenvalues we consider here) is analyzed for random tensors whose entries take values $\pm 1$.} is the ideas of melonic dominance \cite{melon3,melon1,melon2}.  In this letter, we shall focus on a specific quantity of interest where these ideas apply, namely, the largest eigenvalue of a tensor. One important aspect of such largest eigenvalues is that they correspond to eigenvectors that provide the best rank 1 approximation to the tensor \cite{Qirev}. Some properties of the largest eigenvalues, for deterministic tensors, have been discussed in \cite{hyperg,largest}. The key observation is that one can estimate the largest eigenvalue from the rate of growth of a random initial vector under repeated application of a nonlinear map defined by the tensor. This is analogous to the simple power iteration algorithm for finding the largest eigenvalue of a matrix. It turns out that the growth of the norm may be easily analyzed using the melonic dominance techniques, and is effectively determined by a single Feynman diagram. Evaluating this diagram gives an estimate of the largest eigenvalue, which grows as the square root of the number of dimensions $N$, as it does for random matrices. We shall now prove this statement.


\section{The largest eigenvalue of a tensor}

Consider a rank $q+1$ tensor $C_{i_1\cdots i_{q+1}}$ in $N$ spatial directions (each index $i_k$ takes values $1,2,\ldots,N$). Throughout, we shall assume that the tensor entries are real and that the tensor is fully symmetric (any two entries whose indices can be obtained from each other by permutation are equal). The latter property is occasionally referred to as `supersymmetric' in the literature on tensor eigenvalues, but we shall avoid this term to eliminate interference with other meanings of `supersymmetry' in theoretical physics. We shall be interested in the regime of large $N$, which we shall refer to as a {\em large} tensor. This terminology is analogous to calling a matrix `large' when the number of components is large, and has nothing to do with the magnitude of the matrix/tensor entries. In many of our graphical illustrations, we shall fix $q=3$ (a case particularly common from a physicist's perspective), but our analysis is valid for any $q$.

There is no unique generalization of the notion of matrix eigenvalues to tensors, but rather a few different generalizations treated in the literature. In this letter, by `eigenvalues,' we shall mean E-eingenvalues in the terminology of Qi \cite{Qirev}, namely, those numbers $\lambda$ for which the equations
\beq
C_{ii_1\cdots i_q}x_{i_1}\cdots x_{i_q}=\lambda x_i,\qquad x_i x_i=1
\label{eigdef}
\eeq
have solutions. (Throughout, we shall be assuming Einstein's summation conventions, that is, any repeated pair of indices is summed over from 1 to $N$.) Unlike for matrices, one must impose a normalization condition on $x$, since otherwise, rescaling $x$ would have produced a continuum of values of $\lambda$. We note that another definition has received attention in the literature under the name of H-eigenvalues \cite{Qirev}, given by solutions  of
\beq
C_{ii_1\cdots i_q}x_{i_1}\cdots x_{i_q}=\lambda( x_i)^q,
\eeq
where the superscript $q$ denotes an ordinary power of the numerical value of $x_i$.
This definition is not rotationally invariant and less likely to be of interest in physics applications. We shall not investigate it here. (We mention in passing that some powerful methodology to compute the characteristic polynomial whose roots are the H-eigenvalues has been devised in \cite{MSh}.)

We shall be interested in the largest eigenvalue satisfying (\ref{eigdef}). For matrices, to identify the largest eigenvalue, one may apply power iterations. Namely, starting with a random initial vector $x^{(0)}_i$, repeatedly multiply it with the matrix according to $x^{(p+1)}_i=M_{ij}x^{(p)}_j$. For large $p$, the vector will become alligned with the eigenvector of $M$ corresponding to the largest eigenvalue $\lambda_{max}$ and its norm will grow as $\ln(x^{(p)}_ix^{(p)}_i)\sim p\ln(\lambda_{max})$. One can attempt a similar strategy for tensors by choosing a random initial vector $x^{(0)}_i$ and applying iterations of the nonlinear map
\beq
x^{(p+1)}_i=C_{i\,i_1\cdots i_q}x^{(p)}_{i_1}\cdots x^{(p)}_{i_q}.
\label{Cmap}
\eeq
Evidently, if $x^{(0)}_i$ happens to be aligned with an eigenvector (\ref{eigdef}), then the iterations reduce to 
\beq
x^{(p+1)}_i = \lambda\,\,\left(x^{(p)}_kx^{(p)}_k\right)^{\frac{q-1}2}\,\,x^{(p)}_i.
\label{itereigen}
\eeq
Assuming that $x^{(0)}_ix^{(0)}_i=1$, this is solved by
\beq
x^{(p)}_i = \lambda^{[p]_q} x^{(0)}_i,
\label{xpsol}
\eeq
where for compactness of notation, we have introduced the standard $q$-analog of $p$ defined by
\beq
[p]_q\equiv\frac{q^p-1}{q-1}=1+q+\cdots+q^{p-1}.
\label{pq}
\eeq
Note that the growth rate of  the norm of $x^{(p)}$ at large $p$ is tremendous, given by an expo\-nen\-tial-of-an-exponential, so that
\beq
\ln(x^{(p)}_kx^{(p)}_k)\approx 2\ln|\lambda|\,\, \, \frac{q^p-1}{q-1}.
\label{expexp}
\eeq

For matrices, the power iterations will converge to the eigenvector corresponding to the largest eigenvalue, except for initial data of measure zero. For tensors, the situation is vastly more complicated due to the nonlinear character of the map (\ref{Cmap}). As the eigenvectors of $C$ are stationary points of the map (\ref{Cmap}), in the sense that, if reached, the direction of the vector does not change under subsequent iterations as in (\ref{itereigen}), one expects that different eigenvectors possess basins of attraction for the initial data $x^{(0)}_i$. Whether, for a particular tensor $C$, particular initial data  $x^{(0)}_i$ fit into the basin of attraction of the eigenvector corresponding to the largest eigenvalue is expected to depend on the overlap of $x^{(0)}_i$ with that eigenvector, as well as on the distance between the largest eigenvalue and its adjacent smaller eigenvalues. (One can find a systematic discussion of convergence of a related kind of power iterations adapted to H-eigenvalues for the relatively simple case of tensors with nonnegative components in \cite{largest}.)

While, for concrete tensors $C$, the dynamics of the iterated map (\ref{Cmap}) is very complicated, the question that we are addressing here, which is estimating the largest eigenvalue in an ensemble of random tensors, is much simpler. Indeed, we will be effectively averaging both with respect to the tensor components and the initial data. One should expect that, for a given tensor, the norm $x^{(p)}_kx^{(p)}_k$, averaged over the initial data $x^{(0)}_i$ will be dominated by the contribution of the largest eigenvalue. This is because, within the averaging over the initial data, there will be a region where $x^{(0)}_i$ is nearly aligned with the eigenvector corresponding to the largest eigenvalue. Even if this region is small, it will dominate the growth as in (\ref{expexp}), with $\lambda$ replaced by $\lambda_{max}$, since the growth at large $p$ is both tremendously rapid and very sensitive to $\lambda$. Our goal will thus be to use the growth of $x^{(p)}_kx^{(p)}_k$ at large $p$ under the application of the map (\ref{Cmap}) to estimate the largest eigenvalue of $C$ in a Gaussian ensemble to be defined in the next section, and at large $N$.

We remark on some similarity between the questions we address and the {\it spiked tensor model} \cite{CPA, landscapes,talkrivasseau}, which deals with tensors of the form $C_{i_1\cdots i_{q+1}}=v_{i_1}\cdots v_{i_{q+1}} +\text{noise}$. In the ab\-sence of noise, $v_i$ is an eigenvector of $C$. The question of the values of noise for which a recovery of $v$ is possible has been addressed in \cite{CPA}. It was proposed in \cite{talkrivasseau} to consider the flows
\beq
\frac{dx_i}{dt}=C_{i\,i_1\cdots i_q}x_{i_1}\cdots x_{i_q},
\label{Cflow}
\eeq
and to perform an appropriate random tensor average of the trajectories that solve this equation. Similar averages have been performed for other related dynamical systems \cite{meloturb}, in application to deterministic turbulence. A problem with this strategy is that (\ref{Cflow}) is very prone to runaway behaviors, with its trajectories hitting infinity in finite time (this can be immediately seen at $N=1$, where the equation becomes $dx/dt=Cx^q$). If one could make sense of averaging the trajectories of (\ref{Cflow}) over $C$, it could introduce interesting perspectives on stability of nonlinear systems, as a generalization of the classic work \cite{May} on stability on random linear systems. In \cite{landscapes}, the evolution of (\ref{Cflow}) was restricted to a sphere, which prevents blow-up behaviors, and the dynamics of the resulting system was analyzed in the spirit of statistical physics of disordered systems. However, such modifications of (\ref{Cflow}) would make it more difficult to apply the random tensor methodology in the spirit of \cite{meloturb}.

Our approach in this letter will be to focus on the power iterations (\ref{Cmap}) rather than a flow like (\ref{Cflow}), and to address, within this simple setting, the question of the largest tensor eigenvalue. Our approach has a manifest advantage that the finite-time blow-up typical of (\ref{Cflow}) does not occur in the context of (\ref{Cmap}), having been resolved into the super-exponential growth of the norm (\ref{expexp}), and all the random tensor averages at finite $p$ will be well-defined. The diagrammatics associated with random tensor theory will likewise simplify for the power iterations (\ref{Cmap}) as compared to \cite{meloturb}, which is similar in spirit to (\ref{Cflow}).

For a given $p$, it is of course straightforward to apply (\ref{Cmap}) recursively to obtain an expression for $x^{(p)}$ in terms of $C$ and $x^{(0)}$. The index notation, however, becomes extremely impractical for these purposes. The number of copies of $C$ in this expression grows like $[p]_q$ given by (\ref{pq}), and the number of indices necessary, like $[p+1]_q$. It becomes very helpful at this point to switch to the graphic notation for tensor algebra that is sometimes known as the `birdtrack' notation \cite{birdtracks}. In this notation, a rank $k$ tensor is represented as a $k$-valent vertex and its indices are the lines coming out of the vertex. Tensor contractions are simply depicted as connecting the lines coming out of the vertices corresponding to different tensors. (Such graphic notation is particularly common in the tensor network theory \cite{vidal}.) For example, the right-hand side of (\ref{Cmap}) is represented in this language as 
$$
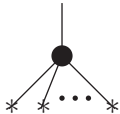
\begin{axopicture}(40,40)
\SetColor{Black}
\Vertex(20,20){4}
\Line(20,20)(20,40)
\Line(20,20)(0,0)
\Text(1,-1)[c]{*}
\Line(20,20)(12,0)
\Text(13,-1)[c]{*}
\Line(20,20)(40,0)
\Text(39,-1)[c]{*}
\Vertex(20,4){1}
\Vertex(25,4){1}
\Vertex(30,4){1}
\end{axopicture}
$$ 
Here, the black vertex denotes $C$ (with $q+1$ lines coming out of it), and the stars represent $x^{(0)}$. The total symmetry of $C$ with respect to index permutations makes it unnecessary to keep track of the order in which the lines exit the black dot. We shall assume $q=3$ in many illustrations below, though the derivations are valid for any $q$.

\begin{figure}[t]
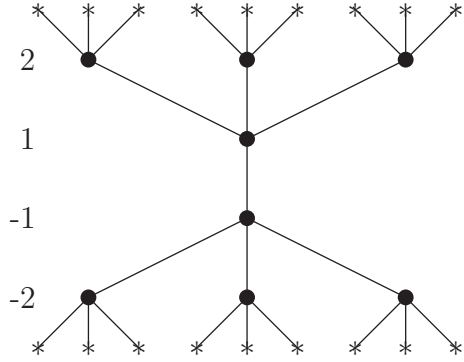

\begin{center}
\begin{axopicture}(160,130)
\SetColor{Black}
\Line(20,20)(0,0)
\Line(20,20)(20,0)
\Line(20,20)(40,0)
\Vertex(20,20){3}
\Line(80,20)(60,0)
\Line(80,20)(80,0)
\Line(80,20)(100,0)
\Vertex(80,20){3}
\Line(140,20)(120,0)
\Line(140,20)(140,0)
\Line(140,20)(160,0)
\Vertex(140,20){3}
\Line(20,20)(80,50)
\Line(80,20)(80,50)
\Line(140,20)(80,50)
\Vertex(80,50){3}
\Line(80,80)(80,50)
\Vertex(80,80){3}
\Line(80,80)(80,110)
\Vertex(80,110){3}
\Line(80,80)(20,110)
\Vertex(20,110){3}
\Line(80,80)(140,110)
\Vertex(140,110){3}
\Line(20,110)(0,130)
\Line(20,110)(20,130)
\Line(20,110)(40,130)
\Line(80,110)(60,130)
\Line(80,110)(80,130)
\Line(80,110)(100,130)
\Line(140,110)(120,130)
\Line(140,110)(140,130)
\Line(140,110)(160,130)
\Text(0,110)[r]{2}
\Text(0,80)[r]{1}
\Text(0,50)[r]{-1}
\Text(0,20)[r]{-2}
\Text(1,-1)[c]{*}
\Text(20,-1)[c]{*}
\Text(39,-1)[c]{*}
\Text(61,-1)[c]{*}
\Text(80,-1)[c]{*}
\Text(99,-1)[c]{*}
\Text(121,-1)[c]{*}
\Text(140,-1)[c]{*}
\Text(159,-1)[c]{*}
\Text(1,127)[c]{*}
\Text(20,127)[c]{*}
\Text(39,127)[c]{*}
\Text(61,127)[c]{*}
\Text(80,127)[c]{*}
\Text(99,127)[c]{*}
\Text(121,127)[c]{*}
\Text(140,127)[c]{*}
\Text(159,127)[c]{*}
\end{axopicture}
\end{center}
\caption{Graphic representation of $x^{(2)}_kx^{(2)}_k$ at $q=3$. The numbers 2, 1, -1, -2 indicate the branching levels. Stars indicate contraction with $x^{(0)}$.}
\label{hourglass}
\end{figure}

Armed with this graphic representation, we can effectively analyze (\ref{Cmap}) at any value of $p$. Namely, $x^{(p)}$ is given by a rooted $(q+1)$-regular tree, where the root corresponds to $x^{(p)}$. There are $p$ levels of regular 1-to-$q$ branching (each black vertex at a given branching level produces $q$ descendant black vertices at the next branching level) and the $q^p$ outgoing leaves are contracted with $q^p$ copies of $x^{(0)}$. If we now want to form $x^{(p)}_kx^{(p)}_k$, we have to take two identical trees for $x^{(p)}$ and connect their roots, yielding an {\em hourglass} diagram, as depicted on fig.~\ref{hourglass}.
The hourglass diagrams appearing from $x^{(p)}_kx^{(p)}_k$ are always obtained by connecting the roots of two identical trees with $p$ levels of branching. For future use, it is convenient to define more generally a $(p,p')$-hourglass diagram, obtained by connecting through the root a tree with $p$ levels of regular branching and a tree with $p'$ levels of regular branching. An hourglass diagram is called {\em balanced} if $p=p'$, and {\em unbalanced} otherwise. As mentioned, the hourglass corresponding  $x^{(p)}_kx^{(p)}_k$ is always balanced, but unbalanced hourglass diagrams will appear in our subsequent considerations. It is convenient to label the branching levels of $(p,p')$-hourglasses by integers in $\{p,p-1,\ldots,1,-1,\ldots,-p'\}$, as in fig.~\ref{hourglass}. The number of edges exiting level $s$ (if moving away from the center of the diagram) is always $q^{|s|}$ due to the completely regular branching at each level.


\section{Gaussian average over the tensor components}
\label{averaging}

We now set out to average $x^{(p)}_kx^{(p)}_k$ represented by a $(p,p)$-hourglass diagram, as in fig.~\ref{hourglass}, over the component of the tensor $C$. The expression for $x^{(p)}_kx^{(p)}_k$, as follows from (\ref{Cmap}), is a homogeneous polynomial of degree $2[p]_q$ in the componenets of $C$.

For the probability distribution of $C$ a Gaussian measure $\exp[-\|C\|^2/2]$, which is rotationally invariant with respect to the standard action of rotations on $C$, and factorizes into independent probability distributions for the components of $C$. We consider the following expectation value of the norm of (\ref{Cmap}), which also contains a more explicit expression for the measure:
\beq
\langle x^{(p)}_kx^{(p)}_k\rangle_{\rule{0mm}{3.2mm}C}=\frac1{\cal N}\int \hspace{-2mm} \prod_{i_1\le i_2\le\cdots\le i_{q+1}} \hspace{-5mm}dC_{i_1\cdots i_{q+1}} \,\exp\left[-\frac12 \sum_{j_1,\ldots, j_{q+1}=1}^N \hspace{-3mm}(C_{j_1\cdots j_{q+1}})^2\,\,\right] x^{(p)}_kx^{(p)}_k.
\label{Gauss}
\eeq
Here, $\cal N$ is the standard normalization factor, given by the same integral without $x^{(p)}_kx^{(p)}_k$.
One must use the total symmetry of $C$ with respect to index permutations to express the argument of the exponential through the independent components $C_{j_1\cdots j_{q+1}}$ with $j_1\le j_2\le\cdots\le j_{q+1}$. This is a direct analog of the Gaussian Orthogonal Ensemble for real symmetric random matrices \cite{mehta}. (Note that the individual entries are independently normally distributed, but the variance of the normal distributions depends on the number of coincident indices in each particular component of $C$.) Gaussian random tensors have recently been treated from a mathematical perspective in \cite{gausstens}. 

For a fixed $C$, in search of its largest eigenvalue, it would have been logical to average over $x^{(0)}$, for instance, average uniformly over $x^{(0)}$ on a unit sphere. Heuristically, in order to converge to the eigenvector corresponding to the largest eigenvalue, $x^{(0)}$ must have some appreciable ($N$-independent) overlap with that eigenvector (when $N$ is large). The fraction of the volume of the unit sphere where the overlap with a given vector is a fixed number will be exponentially suppressed at large $N$ as $e^{-\eta N}$, where $\eta$ is some numerical factor (this exponential suppression underlies the predominance of the `equatorial region' with its `glassy' dynamics in \cite{landscapes}). The exponential suppression is, however, negligible compared to the superexponential growth given by (\ref{expexp}), and the large $p$ asymptotics of $x^{(p)}_kx^{(p)}_k$, averaged over the direction of $x^{(0)}$ is expected to be dominated by the largest eigenvalue.

If one averages over $C$ as in (\ref{Gauss}), the averaging over the direction of $x^{(0)}$ becomes irrelevant and will not change the expectation value (\ref{Gauss}), since the ensemble given by (\ref{Gauss}) is rotationally invariant. One can thus perform all computations for a fixed unit vector $x^{(0)}$, while an averaging over its direction is already implicit in the averaging over $C$ in (\ref{Gauss}), and the remarks of the previous paragraph apply.

Our subsequent discussion of the average of $C$ within the Gaussian ensemble (\ref{Gauss}) is completely standard in random tensor theory \cite{melon1,melon2,melon3}, though the particular simplicity of our diagrammatics allows for an elementary self-contained presentation, which is what will be given here. By Wick's theorem \cite{ZJ}, evaluating averages of any polynomial expression in $C$, as in (\ref{Gauss}), reduces to summing over all possible pairings of $C$, whereupon, each doublet of paired $C$'s is replaced by the corresponding covariance matrix, or the `propagator,' using the language of Feynman diagrams, given by
\beq
\langle C_{i_1\cdots i_{q+1}}C_{k_1\cdots k_{q+1}}\rangle_{\rule{0mm}{3.2mm}C}=\frac1{(q+1)!}\sum_{\sigma({\bf k})} \de_{i_1\sigma_1({\bf k})} \de_{i_2\sigma_2({\bf k})}\cdots \de_{i_{q+1}\sigma_{q+1}({\bf k})}.
\label{CC}
\eeq
Here, the summation is over all permutations $\sigma$ of the set ${\bf k}\equiv\{k_1,k_2,\ldots,k_{q+1}\}$ rearranging it into $\{\sigma_1({\bf k}),\sigma_2({\bf k}),\ldots,\sigma_{q+1}({\bf k})\}$. (Note that all the $(q+1)!$ permutations must be counted in the above summation, even if a particular permutation leaves a particular set of indices invariant.)

\begin{figure}[t]
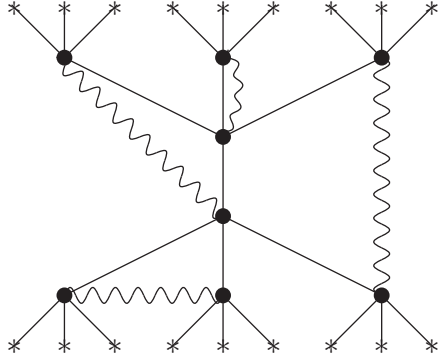

\begin{center}
\begin{axopicture}(160,130)
\SetColor{Black}
\Line(20,20)(0,0)
\Line(20,20)(20,0)
\Line(20,20)(40,0)
\Vertex(20,20){3}
\Line(80,20)(60,0)
\Line(80,20)(80,0)
\Line(80,20)(100,0)
\Vertex(80,20){3}
\Line(140,20)(120,0)
\Line(140,20)(140,0)
\Line(140,20)(160,0)
\Vertex(140,20){3}
\Line(20,20)(80,50)
\Line(80,20)(80,50)
\Line(140,20)(80,50)
\Vertex(80,50){3}
\Line(80,80)(80,50)
\Vertex(80,80){3}
\Line(80,80)(80,110)
\Vertex(80,110){3}
\Line(80,80)(20,110)
\Vertex(20,110){3}
\Line(80,80)(140,110)
\Vertex(140,110){3}
\Line(20,110)(0,130)
\Line(20,110)(20,130)
\Line(20,110)(40,130)
\Line(80,110)(60,130)
\Line(80,110)(80,130)
\Line(80,110)(100,130)
\Line(140,110)(120,130)
\Line(140,110)(140,130)
\Line(140,110)(160,130)
\Photon(20,20)(80,20){3}{7}
\Photon(80,50)(17,110){3}{10}
\Photon(140,20)(140,110){3}{10}
\PhotonArc(56,98)(30,328,30){1.5}{4}
\Text(1,-1)[c]{*}
\Text(20,-1)[c]{*}
\Text(39,-1)[c]{*}
\Text(61,-1)[c]{*}
\Text(80,-1)[c]{*}
\Text(99,-1)[c]{*}
\Text(121,-1)[c]{*}
\Text(140,-1)[c]{*}
\Text(159,-1)[c]{*}
\Text(1,127)[c]{*}
\Text(20,127)[c]{*}
\Text(39,127)[c]{*}
\Text(61,127)[c]{*}
\Text(80,127)[c]{*}
\Text(99,127)[c]{*}
\Text(121,127)[c]{*}
\Text(140,127)[c]{*}
\Text(159,127)[c]{*}
\end{axopicture}
\end{center}
\caption{One particular pairing of the black vertices with wavy lines resulting from performing a Gaussian average in fig.~\ref{hourglass}.}
\label{Ccontr}
\end{figure}

As the number and complexity of tensor contractions increases after the averaging over $C$, the graphic notation becomes even more indispensable. The pairings of $C$ mandated by Wick's theorem are represented by additional wavy lines connecting the black vertices corresponding to $C$. Each black vertex must belong to exactly one such wavy line. An example of such contraction, derived from fig.~\ref{hourglass}, is given in fig.~\ref{Ccontr}.
After the pairings have been implemented, one evaluates the diagram by substituting (\ref{CC}) for each pairing, which simply amounts to implementing the following graphical operation\vspace{4mm}
\beq
\begin{axopicture}(45,0)
\SetColor{Black}
\Vertex(20,-10){4}
\Line(20,-10)(0,-30)
\Line(20,-10)(12,-30)
\Line(20,-10)(40,-30)
\Vertex(20,-26){1}
\Vertex(25,-26){1}
\Vertex(30,-26){1}
\Photon(20,-10)(20,10){2}{3}
\Vertex(20,10){4}
\Line(20,10)(0,30)
\Line(20,10)(12,30)
\Line(20,10)(40,30)
\Vertex(20,26){1}
\Vertex(25,26){1}
\Vertex(30,26){1}
\end{axopicture}
=\frac{1}{(q+1)!}\,\,\Bigg\{\,\,\,\,
\begin{axopicture}(30,0)
\Line(5,-7)(5,13)
\Line(10,-7)(10,13)
\Vertex(15,3){1}
\Vertex(20,3){1}
\Vertex(25,3){1}
\Line(30,-7)(30,13)
\end{axopicture}
\,\,\,\,\,+\,\,\,\,
\begin{axopicture}(30,0)
\Line(5,-7)(10,13)
\Line(10,-7)(5,13)
\Vertex(15,3){1}
\Vertex(20,3){1}
\Vertex(25,3){1}
\Line(30,-7)(30,13)
\end{axopicture}
\,\,\,\,\,+\,\,\,\,\ldots\,\,\,\,\Bigg\},
\label{Cprop}
\eeq\vspace{5mm}

\noindent where the sum on the right hand side is over all the $(q+1)!$ ways to connect the $q+1$ endpoints at the top to the $q+1$ endpoints at the bottom. After this substitution has been made, all black vertices and all wavy lines disappear, and one is left with products of Kronecker symbols from (\ref{CC}) contracted along ordinary (non-wavy) lines. These lines can either end on stars, corresponding to contraction with $x^{(0)}$, which simply gives $x^{(0)}_kx^{(0)}_k=1$, or they can form a loop, in which case one gets from that particular line $\de_{kk}=N$. Thus, after the evaluation of the Gaussian average over $C$ has been implemented with the graphical operations we have described, each resulting diagram depends on $N$ as simply $N$ to the power of the number of loops.

In practice, when one computes (\ref{Gauss}) at fixed $p$, one obtains a polynomial in $N$ whose degree is bounded by the largest number of loops attainable at that value of $p$. Since we are interested in large $N$, only the highest order term of this polynomial is relevant for us. Our strategy for the rest of the treatment will be thus to bound the largest possible number of loops in the diagrammatic decomposition of (\ref{Gauss}), to identify the corresponding diagrams, and to compute their contribution to (\ref{Gauss}) given by a single specific power of $N$.


\section{The dominant diagram}

We shall now identify the diagrams arising from the $C$-averaging of the previous section and yielding the highest power of $N$. The analysis is a particularly simple realization of the idea of `melonic dominance' typical of the random tensor theory \cite{melon1,melon2,melon3}.

\subsection{Faces and chains}

As explained at the end of the previous section, substitution of (\ref{Cprop}) into diagrams of the sort depicted on fig.~\ref{Ccontr} arising from the Feynman graph expansion of (\ref{Gauss}) converts these diagrams into collections of lines ending on stars, contributing no powers of $N$, and loops, contributing a power of $N$ each. Below, we shall refer to the lines as {\em chains} and to the loops as {\em faces}.

For the following analysis, we introduce the notion of {\em doubled vertices}. Namely, in (\ref{Cprop}), one can think of the graphic element {}
\begin{axopicture}(20,0)
\Vertex(0,3.5){4}
\Photon(0,3.5)(20,3.5){2}{3}
\Vertex(20,3.5){4}
\end{axopicture}
\hspace{1.7mm} as a new kind of vertex, to which $2(q+1)$ lines are attached. In other words, we replace the right-hand side of (\ref{Cprop}) by the following graphic representation:\vspace{4mm}
\beq
\begin{axopicture}(45,0)
\SetColor{Green}
\Line(20,-10)(0,-30)
\Line(20,-10)(12,-30)
\Line(20,-10)(40,-30)
\Vertex(20,-26){1}
\Vertex(25,-26){1}
\Vertex(30,-26){1}
\SetColor{Red}
\Line(20,10)(0,30)
\Line(20,10)(12,30)
\Line(20,10)(40,30)
\Vertex(20,26){1}
\Vertex(25,26){1}
\Vertex(30,26){1}
\SetColor{Black}
\Vertex(20,-10){4}
\Photon(20,-10)(20,10){2}{3}
\Vertex(20,10){4}
\end{axopicture}
=\,\,\Bigg\{\hspace{1cm}
\begin{axopicture}(30,0)
\Arc(0,3)(15,10,24)
\Arc(0,3)(15,38,52)
\Arc(0,3)(15,66,80)
\SetColor{Red}
\Line(-2,5)(-2,25)
\Line(-2,1)(-22,1)
\Line(2,1)(2,-19)
\SetColor{Green}
\Line(2,1)(22,1)
\Line(-2,1)(-2,-19)
\Line(-2,5)(-22,5)
\end{axopicture}
\Bigg\}_{\text{symm}},
\label{doubledvertex}
\eeq\vspace{5mm}

\noindent where the red and green color is purely to indicate which segment comes from which side of the original diagram and does not affect computations, and ``symm'' indicates a total symmetrization with respect to the ways the green segments are paired with the red ones, as on the right-hand side of (\ref{Cprop}). The new doubled vertex, of course, does not affect evaluation of the diagrams and can be simply dissolved to recover the right-hand side of (\ref{Cprop}). It is, however, useful for bookkeeping of the graphs. In particular, each face and each chain touches a particular sequence of such doubled vertices. One can then define the {\em perimeter} of a chain or a face, according to the number of corners around doubled vertices that it touches along its way. We will use $F_l$ with $l\ge 1$ to denote the number of faces of perimeter $l$, and similarly, $C_l$ for chains of perimeter $l$.

With these preliminaries, we can derive an upper bound on the number of faces, and hence the power of $N$, arising via application of $C$-averaging to $x^{(p)}_kx^{(p)}_k$ represented by a $(p,p)$-hourglass diagram, as exemplified by fig.~\ref{hourglass}.  This estimate is the central point of our computations.

For future use, it is wise to consider the following more general problem: instead of a $(p,p)$-hourglass diagram, take a collection of $K$ arbitrary trees with $2L$ univalent vertices, depicted as stars, $2V$ vertices of valence $(q+1)$, depicted as black dots, and E edges. By elementary properties of trees,
\beq
E=2L+2V-K.
\label{edgeexp}
\eeq
We now apply the $C$-averaging, as in section \ref{averaging}, which results in contraction of the $(q+1)$-valent vertices and formation of $V$ doubled vertices. The new graph consists of $2L$ univalent vertices, $V$ doubled vertices, and $E$ edges. The number of connected components can only decrease through contraction of $(q+1)$-valent vertices into doubled vertices, so that the new number of connected components $\tilde K$ satisfies
\beq
\tilde K\le K.
\eeq
By straightforward counting of edges along the faces and the chains, one has
\beq
\sum_l l F_l+\sum_l (l+1) C_l=E,
\label{edgecount}
\eeq
as a face of perimeter $l$ consists of $l$ edges, while a chain of perimeter $l$ consists of $(l+1)$ edges. We can think of forming this graph by starting with $V$ disconnected doubled vertices and adding the faces and chains one-by-one. Each face or chain of perimeter $l$ decreases the number of disconnected components by at most $l-1$. Since one should end up with $\tilde K$ disconnected components,
\beq
\sum_l (l-1) F_l+\sum_l (l-1) C_l\ge V-\tilde K.
\label{concomp}
\eeq
Subtracting (\ref{concomp}) from (\ref{edgecount}), we get
\beq
\sum_l  F_l+ 2 \sum_l C_l\le E-V+\tilde K\le E-V+ K.
\label{KKtilde}
\eeq
We note that $\sum_l C_l$ is the total number of chains, which is simply $L$, since each chain must connect exactly two stars. Then, taking into account (\ref{edgeexp}), one simply concludes that
\beq
\sum_l  F_l\le V,
\label{Flbound}
\eeq
so the total number of faces is no greater than $V$ and the diagram grows at most as $N^V$. Note that (\ref{KKtilde}) can only be saturated if $\tilde K=K$, which means that, at the dominant power of $N$, wavy lines never connect different disconnected components of the original collection of $K$ trees.

If we apply the above bound to $(p,p)$-hourglass diagrams, as in fig.~\ref{hourglass}, the number of black vertices $2V$ is simply $2[p]_q$ as given by (\ref{pq}), and hence one  anticipates an estimate
\beq
\langle x^{(p)}_kx^{(p)}_k\rangle_{\rule{0mm}{3.2mm}C}\,\,\sim \,\,N^{[p]_q}.
\label{normCprop}
\eeq
This is the central result of our treatment, though a few details remain to be filled in, which is the purpose of the rest of the letter. Indeed, one must show that diagrams with this particular scaling exist, and better still, to characterize the full set of such diagrams, and supply the numerical coefficient on the right-hand side of  (\ref{normCprop}).

\subsection{Recursive reduction of faces of perimeter 1}

We now return to the setup described above (\ref{edgeexp}) and assume that the bound (\ref{Flbound}) is saturated, so that $\sum_l F_l=V$, in order to explore the properties of such dominant diagrams. Evidently,
\beq
F_1\ge 2\sum_lF_l-\sum_l lF_l= 2V-E+\sum_l (l+1)C_l,
\eeq
where we have used (\ref{edgecount}). We then write $\sum_l (l+1)C_l\ge2\sum_l C_l=2L$. Hence, using (\ref{edgeexp}),
\beq
F_1\ge K\ge 1.
\eeq
Thus, any dominant diagram must contain at least one face of perimeter 1. This situation is typical of the dominant `melonic' diagrams of random tensor theory \cite{melon1,melon2,melon3,meloturb}.

We then take a dominant diagram and zoom in on a particular face of perimeter 1, which must necessarily be present. In the part of the diagram surrounding this face of perimeter 1, we can apply (\ref{Cprop}) as follows\vspace{6mm}
\beq
\begin{axopicture}(45,0)
\SetColor{Black}
\Vertex(20,-10){4}
\Line(20,-10)(0,-30)
\Line(20,-10)(12,-30)
\Line(20,-10)(40,-30)
\Vertex(20,-26){1}
\Vertex(25,-26){1}
\Vertex(30,-26){1}
\PhotonArc(20,0)(10,90,270){2}{4}
\Line(20,-10)(20,10)
\Vertex(20,10){4}
\Line(20,10)(0,30)
\Line(20,10)(12,30)
\Line(20,10)(40,30)
\Vertex(20,26){1}
\Vertex(25,26){1}
\Vertex(30,26){1}
\end{axopicture}
=\Bigg\{\,\,\frac{1}{q+1}\hspace{7mm}
\begin{axopicture}(30,0)
\Arc(-8,3)(10,0,360)
\Line(5,-7)(5,13)
\Line(10,-7)(10,13)
\Vertex(15,3){1}
\Vertex(20,3){1}
\Vertex(25,3){1}
\Line(30,-7)(30,13)
\end{axopicture}
\,\,\,\,\,+\,\,\frac{q}{q+1}
\begin{axopicture}(30,0)
\Line(5,-7)(5,13)
\Line(10,-7)(10,13)
\Vertex(15,3){1}
\Vertex(20,3){1}
\Vertex(25,3){1}
\Line(30,-7)(30,13)
\end{axopicture}\hspace{1mm}\,\,\Bigg\}_{\text{symm}}\,\,\,,
\label{face1red}
\eeq\vspace{5mm}

\noindent where the first term on the right-hand side comes from terms in (\ref{Cprop}) that connect the endpoints of the straight vertical line between the two black vertices, resulting in a loop, with the coefficient given by the total number of ways to connect the remaining $2q$ straight lines with each other, which is $q!$, divided by $(q+1)!$ on the right-hand side of (\ref{Cprop}). The second term comes from all the remaining contributions to (\ref{Cprop}). In analogy to the previous formulas, ``symm'' can be understood as summation over all the $q!$ permutations of the $q$ lower endpoints, followed with division by $q!$ (this ``symm'' operation can  in practice be ignored for the evaluation of the dominant diagram that we will recover at the end, but we shall keep it explicit for now).

\begin{figure}[t]
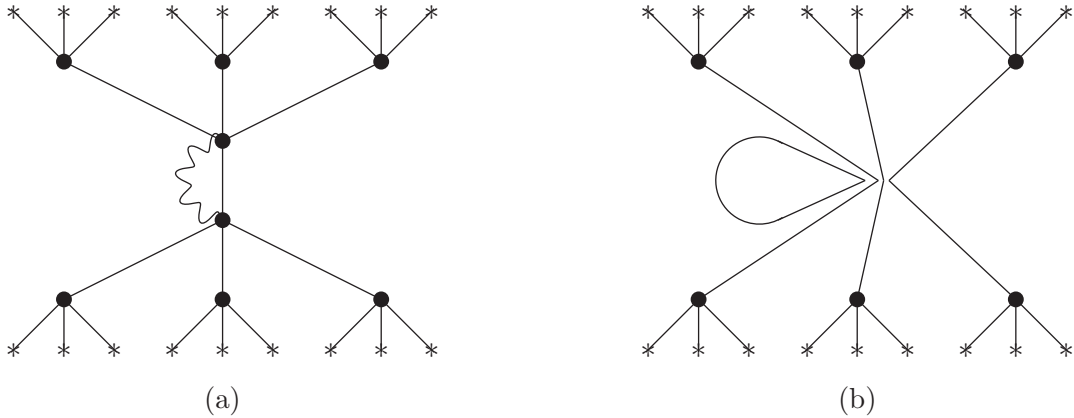

\centering
    \begin{subfigure}[b]{0.5\textwidth}
\centering
\begin{axopicture}(160,130)
\SetColor{Black}
\Line(20,20)(0,0)
\Line(20,20)(20,0)
\Line(20,20)(40,0)
\Vertex(20,20){3}
\Line(80,20)(60,0)
\Line(80,20)(80,0)
\Line(80,20)(100,0)
\Vertex(80,20){3}
\Line(140,20)(120,0)
\Line(140,20)(140,0)
\Line(140,20)(160,0)
\Vertex(140,20){3}
\Line(20,20)(80,50)
\Line(80,20)(80,50)
\Line(140,20)(80,50)
\Vertex(80,50){3}
\Line(80,80)(80,50)
\Vertex(80,80){3}
\Line(80,80)(80,110)
\Vertex(80,110){3}
\Line(80,80)(20,110)
\Vertex(20,110){3}
\Line(80,80)(140,110)
\Vertex(140,110){3}
\Line(20,110)(0,130)
\Line(20,110)(20,130)
\Line(20,110)(40,130)
\Line(80,110)(60,130)
\Line(80,110)(80,130)
\Line(80,110)(100,130)
\Line(140,110)(120,130)
\Line(140,110)(140,130)
\Line(140,110)(160,130)
\PhotonArc(80,65)(15,90,270){3}{5}
\Text(1,-1)[c]{*}
\Text(20,-1)[c]{*}
\Text(39,-1)[c]{*}
\Text(61,-1)[c]{*}
\Text(80,-1)[c]{*}
\Text(99,-1)[c]{*}
\Text(121,-1)[c]{*}
\Text(140,-1)[c]{*}
\Text(159,-1)[c]{*}
\Text(1,127)[c]{*}
\Text(20,127)[c]{*}
\Text(39,127)[c]{*}
\Text(61,127)[c]{*}
\Text(80,127)[c]{*}
\Text(99,127)[c]{*}
\Text(121,127)[c]{*}
\Text(140,127)[c]{*}
\Text(159,127)[c]{*}
\end{axopicture}\vspace{2mm}
\caption{}
 \end{subfigure}%
    ~ 
    \begin{subfigure}[b]{0.5\textwidth}
\centering
\begin{axopicture}(160,130)
\SetColor{Black}
\Line(20,20)(0,0)
\Line(20,20)(20,0)
\Line(20,20)(40,0)
\Vertex(20,20){3}
\Line(80,20)(60,0)
\Line(80,20)(80,0)
\Line(80,20)(100,0)
\Vertex(80,20){3}
\Line(140,20)(120,0)
\Line(140,20)(140,0)
\Line(140,20)(160,0)
\Vertex(140,20){3}
\Line(20,20)(88,65)
\Line(80,20)(90,65)
\Line(140,20)(92,65)
\Line(83,65)(50,80)
\Line(83,65)(50,50)
\Arc(43,65)(16.5,58,302)
\Line(90,65)(80,110)
\Vertex(80,110){3}
\Line(88,65)(20,110)
\Vertex(20,110){3}
\Line(92,65)(140,110)
\Vertex(140,110){3}
\Line(20,110)(0,130)
\Line(20,110)(20,130)
\Line(20,110)(40,130)
\Line(80,110)(60,130)
\Line(80,110)(80,130)
\Line(80,110)(100,130)
\Line(140,110)(120,130)
\Line(140,110)(140,130)
\Line(140,110)(160,130)
\Text(1,-1)[c]{*}
\Text(20,-1)[c]{*}
\Text(39,-1)[c]{*}
\Text(61,-1)[c]{*}
\Text(80,-1)[c]{*}
\Text(99,-1)[c]{*}
\Text(121,-1)[c]{*}
\Text(140,-1)[c]{*}
\Text(159,-1)[c]{*}
\Text(1,127)[c]{*}
\Text(20,127)[c]{*}
\Text(39,127)[c]{*}
\Text(61,127)[c]{*}
\Text(80,127)[c]{*}
\Text(99,127)[c]{*}
\Text(121,127)[c]{*}
\Text(140,127)[c]{*}
\Text(159,127)[c]{*}
\end{axopicture}\vspace{2mm}
\caption{}
    \end{subfigure}
    \caption{The left-hand side (a) and the first term of the right-hand side (b) of (\ref{face1red}) in application to a particular possible face of perimeter 1, visible as a loop in (b), resulting from contraction of black dots in fig.~\ref{hourglass}.}
\label{face1}
\end{figure}

Of course, (\ref{face1red}) must be graphically embedded into the full diagram to which the face of perimeter 1 belongs. This results in having the upper and lower enpoints on the right-hand side of (\ref{face1red}) connected to trees made of straigh lines, black vertices and stars (an example is given in fig.~\ref{face1}). Thereafter, the computation has to be continued by contracting the remaining black vertices. Note that after embedding (\ref{face1red}) into a bigger diagram, the diagrams corresponding to the two terms on the right-hand side of (\ref{face1red}) are exactly the same, except for the extra loop contained in the first term, which corresponds to an extra power of N and means that the second term is necessarily subleading. Thus, at leading order in $N$, one simply has\vspace{6mm}
\beq
\begin{axopicture}(45,0)
\SetColor{Black}
\Vertex(20,-10){4}
\Line(20,-10)(0,-30)
\Line(20,-10)(12,-30)
\Line(20,-10)(40,-30)
\Vertex(20,-26){1}
\Vertex(25,-26){1}
\Vertex(30,-26){1}
\PhotonArc(20,0)(10,90,270){2}{4}
\Line(20,-10)(20,10)
\Vertex(20,10){4}
\Line(20,10)(0,30)
\Line(20,10)(12,30)
\Line(20,10)(40,30)
\Vertex(20,26){1}
\Vertex(25,26){1}
\Vertex(30,26){1}
\end{axopicture}
=\Bigg\{\,\,\frac{N}{q+1}
\begin{axopicture}(30,0)
\Line(5,-7)(5,13)
\Line(10,-7)(10,13)
\Vertex(15,3){1}
\Vertex(20,3){1}
\Vertex(25,3){1}
\Line(30,-7)(30,13)
\end{axopicture}\hspace{2mm}\,\,\Bigg\}_{\text{symm}}\,\,\,.
\label{face1reddom}
\eeq\vspace{5mm}

\noindent After embedding the above substitution in the full diagram, one obtains a collection of $q$ disconnected trees whose black vertices are to be contracted with wavy lines. As explained under (\ref{Flbound}), if the contractions are to lead to the dominant power of $N$, they cannot reconnect the different disconnected components. Therefore, for the rest of the computation, the dominant contribution factorizes into a product of independent contributions from each of the $q$ trees. 
We shall now recursively apply this procedure of reducing faces of perimeter 1, and identify the dominant diagram.

\subsection{Analysis of the dominant diagram}

As each dominant diagram must contain a face of perimeter 1, one applies (\ref{face1reddom}) to reduce this face, as a result of which the diagram splits into a collection of $q$ independent trees. Each of these trees must contain a face of perimeter 1 in order to generate the dominant power of $N$, and hence the process will continue until there are no black vertices left to contract. If, in this process, one of the disconnected components consists of a single black vertex, it evidently cannot be contracted with other black vertices to form a face of perimeter 1, and hence the contraction pattern that has induced this situation does not generate the dominant power of $N$ and must be discarded. This recursive consideration allows for an easy identification of the dominant diagram. We start by proving two auxiliary statements:\vspace{3mm}

\noindent (1) {\em  If one starts with a balanced $(p,p)$-hourglass diagram and places the face of perimeter 1 anywhere else than between the branching levels 1 and -1 (see fig.~\ref{hourglass}), the collection of $q$ trees resulting from the reduction of this face of perimeter 1 will contain an unbalanced $(p_1,p_2)$-hourglass diagram ($p_1\ne p_2$).}\vspace{2mm}

\noindent Indeed, assume that the face of perimeter 1 is placed between branching levels $r$ and $r+1$ (due to symmetry, one is allowed to take $r>0$). Then, the part of the diagram surrounding this face of perimeter 1 will look like
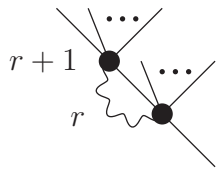
\vspace{8mm}
$$
\begin{axopicture}(45,0)
\SetColor{Black}
\Vertex(40,-10){4}
\Line(40,-10)(32,10)
\Line(40,-10)(60,10)
\Vertex(40,6){1}
\Vertex(45,6){1}
\Vertex(50,6){1}
\PhotonArc(30,0)(13,135,315){2}{4.5}
\Line(40,-10)(20,10)
\Vertex(20,10){4}
\Line(40,-10)(60,-30)
\Line(20,10)(0,30)
\Line(20,10)(12,30)
\Line(20,10)(40,30)
\Vertex(20,26){1}
\Vertex(25,26){1}
\Vertex(30,26){1}
\Text(-5,10){$r+1$}
\Text(8,-12){$r$}
\end{axopicture}
$$\vspace{5mm}

\noindent Because the original $(p,p)$-hourglass is made of two $q$-regular trees with $p$ levels of regular branching, all upward segments exiting the branching level $r+1$ in the picture above are in fact $q$-regular trees with $p-r-1$ levels of regular branching, and all upward segments exiting the branching level $r$ 
are $q$-regular trees with $p-r$ levels of regular branching. Thus, pairing these segments via application of (\ref{Cprop}) will generate $(p\!-\!r\!-\!1,p\!-\!r)$-hourglass diagrams, necessarily unbalanced. (The downward segment exiting the branching level $r$ does not in general generate an hourglass diagram via application of (\ref{Cprop}), but a more general tree, but that is irrelevant for our argument.) Note that if, contrary to our assumption, the face of perimeter 1 is placed between branching levels 1 and -1, as in fig.~\ref{face1}a, one ends up with a collection of $q$ balanced $(p\!-\!1,p\!-\!1)$-hourglass diagrams.\vspace{3mm}

\noindent (2) {\em  If one starts with an unbalanced $(p,p')$-hourglass diagram ($p\ne p'$) and reduces its face of perimeter 1, the resulting collection of $q$ trees always contains an unbalanced hourglass diagram.}\vspace{2mm}

\noindent Indeed, if the face of perimeter 1 is placed anywhere other than between branching levels 1 and -1, the proof is as above, and if it is placed between the branching levels 1 and -1, application of (\ref{Cprop}) results in a collection of $q$ copies of the $(p\!-\!1,p'\!-\!1)$-hourglass diagram, evidently unbalanced.\vspace{3mm}

Statement (2) means that, in order to obtain the dominant power of $N$, in the process of recursive reduction of faces of perimeter 1, unbalanced hourglass diagrams should never emerge. If they ever emerge, then in any subsequent step of recursive reduction of faces of perimeter 1, unbalanced hourglass diagrams must be present. Since the number of black vertices decreases by 2 at each step, one must necessarily end up with at least one $(1,0)$-hourglass diagram, i.e., an isolated black vertes. But such a vertex can no longer be contracted to form a face of perimeter 1, and hence the diagram under consideration is subdominant and can be discarded.

\begin{figure}[t]
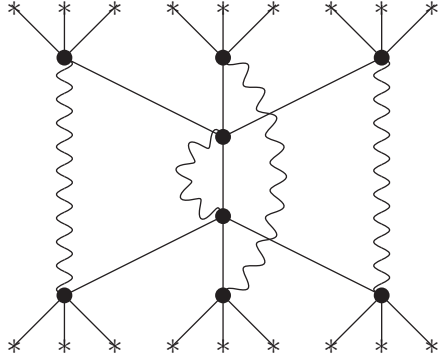

\centering
\begin{axopicture}(160,130)
\SetColor{Black}
\Line(20,20)(0,0)
\Line(20,20)(20,0)
\Line(20,20)(40,0)
\Vertex(20,20){3}
\Line(80,20)(60,0)
\Line(80,20)(80,0)
\Line(80,20)(100,0)
\Vertex(80,20){3}
\Line(140,20)(120,0)
\Line(140,20)(140,0)
\Line(140,20)(160,0)
\Vertex(140,20){3}
\Line(20,20)(80,50)
\Line(80,20)(80,50)
\Line(140,20)(80,50)
\Vertex(80,50){3}
\Line(80,80)(80,50)
\Vertex(80,80){3}
\Line(80,80)(80,110)
\Vertex(80,110){3}
\Line(80,80)(20,110)
\Vertex(20,110){3}
\Line(80,80)(140,110)
\Vertex(140,110){3}
\Line(20,110)(0,130)
\Line(20,110)(20,130)
\Line(20,110)(40,130)
\Line(80,110)(60,130)
\Line(80,110)(80,130)
\Line(80,110)(100,130)
\Line(140,110)(120,130)
\Line(140,110)(140,130)
\Line(140,110)(160,130)
\PhotonArc(80,65)(15,90,270){3}{5}
\PhotonArc(35,65)(66,-45,45){-3}{9.5}
\Photon(140,20)(140,110){3}{10}
\Photon(20,20)(20,110){3}{10}
\Text(1,-1)[c]{*}
\Text(20,-1)[c]{*}
\Text(39,-1)[c]{*}
\Text(61,-1)[c]{*}
\Text(80,-1)[c]{*}
\Text(99,-1)[c]{*}
\Text(121,-1)[c]{*}
\Text(140,-1)[c]{*}
\Text(159,-1)[c]{*}
\Text(1,127)[c]{*}
\Text(20,127)[c]{*}
\Text(39,127)[c]{*}
\Text(61,127)[c]{*}
\Text(80,127)[c]{*}
\Text(99,127)[c]{*}
\Text(121,127)[c]{*}
\Text(140,127)[c]{*}
\Text(159,127)[c]{*}
\end{axopicture}\vspace{3mm}
\caption{The dominant contraction pattern corresponding to fig.~\ref{hourglass}.}
\label{melon}
\end{figure}

With this in mind, starting with a $(p,p)$-hourglass diagram representing (\ref{Gauss}), we must contract the two black vertices at branching levels 1 and -1, as in fig.~\ref{face1}a, since if we did not, the face of perimeter 1 would be present at a different branching level, and its reduction would have produced an unbalanced hourglass diagram per statement (1) above. Hence, branching levels 1 and -1 are contracted as in fig.~\ref{face1}a, and reduction of that face of perimeter 1 (fig.~\ref{face1}b) produces $q$ balanced $(p\!-\!1,p\!-\!1)$-hourglass diagrams. The argument is then recursively repeated for each of these $(p\!-\!1,p\!-\!1)$-hourglass diagrams, until there are no black vertices left. The result is an evident contraction pattern, exemplified by fig.~\ref{melon}, which is the only contraction leading to the dominant power of $N$. 
Evaluation of this dominant diagram proceeds straightforwardly by recursive application of (\ref{face1reddom}), starting with fig.~\ref{face1}b, which yields, at large $N$,
\beq
\langle x^{(p)}_kx^{(p)}_k\rangle_{\rule{0mm}{3.2mm}C}=\left(\frac{N}{q+1}\right)^{[p]_q}.
\label{normN}
\eeq

The $q$-dependent combinatorial factor in (\ref{normN}) can be more systematically understood as follows. Consider a $(p,p)$-hourglass diagram representing (\ref{Gauss}) and implement all possible contractions of the black vertices. According to the recursive analysis above, the dominant diagrams are defined by the following two rules:
\begin{enumerate}
\item Black vertices at branching level $r$ can only be paired with black vertices at branching level $-r$.
\item If a black vertex A at branching level $r$ is paired with a black vertex B at branching level $-r$, the $q$ descendants of A at branching level $r+1$ can only be paired with descendants of B at branching level $-(r+1)$.
\end{enumerate}
The number of such diagrams is $(q!)^{[p]_q}$ by elementary combinatorial considerations. Because of the total symmetry of $C$, the straight lines can be freely permuted around the black vertex where they originate, and hence all the dominant diagrams described above are equivalent. The permutations can be used to reorder the black vertices within the horizontal rows so that each black vertex at branching level $r$ is paired with the black vertex directly below it at branching level $-r$, as in fig.~\ref{melon}. We first reduce the face of perimeter 1 located between branching levels 1 and -1. When substituting (\ref{face1reddom}), only the first term within the sum over all permutations on the right-hand side (without any line crossings), will generate the maximal power of $N$. The result is $N/((q+1)q!)$, where $q!$ comes from the symmetrization operation in (\ref{face1reddom}), times $q$ copies of the $(p\!-\!1,p\!-\!1)$-hourglass diagram with `vertical' contractions of black vertices, as in fig.~\ref{melon}. Repeating this procedure recursively, the diagram with `vertical' contractions is evaluated as $(N/(q+1)!)^{[p]_q}$. Since there are $(q!)^{[p]_q}$ diagrams total with this value, the result is (\ref{normN}).
Then, from (\ref{expexp}) and (\ref{normN}), one gets the largest eigenvalue estimate
\beq
|\lambda_{max}|\approx \sqrt\frac{N}{q+1}.
\label{lambdamax}
\eeq

\begin{figure}[t]
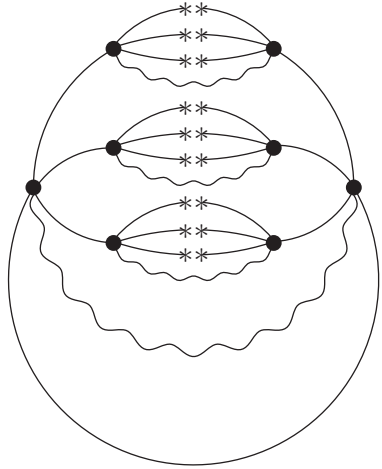

\centering
\begin{axopicture}(160,170)
\SetColor{Black}
\Arc(70,70)(70,150,30)
\Vertex(9.4,105){3}
\Vertex(130.6,105){3}
\PhotonArc(70,103)(60,180,0){2}{10.3}
\Arc(70,105)(60.6,0,60)
\Arc(70,105)(60.6,120,180)
\Vertex(39.7,157.5){3}
\Vertex(100.3,157.5){3}
\Vertex(39.7,120){3}
\Vertex(100.3,120){3}
\Vertex(39.7,84){3}
\Vertex(100.3,84){3}
\Arc(41,119)(35,207,270)
\Arc(98.5,120.5)(36,270,333)
\Arc(39,85)(35,90,150)
\Arc(101,85)(36,30,90)
\Arc(70,134)(39,41,85)
\Arc(70,134)(39,95,139)
\Text(67,170.7)[c]{*}
\Text(73,170.7)[c]{*}
\Arc(70,68)(95,70,88)
\Arc(70,68)(95,92,110)
\Text(67,161)[c]{*}
\Text(73,161)[c]{*}
\Arc(70,248)(95,250,268)
\Arc(70,248)(95,272,290)
\Text(67,151)[c]{*}
\Text(73,151)[c]{*}
\PhotonArc(70,192)(48,227,313){-1}{6.5}
\Arc(70,96.5)(39,41,85)
\Arc(70,96.5)(39,95,139)
\Text(67,133.2)[c]{*}
\Text(73,133.2)[c]{*}
\Arc(70,30.5)(95,70,88)
\Arc(70,30.5)(95,92,110)
\Text(67,123.2)[c]{*}
\Text(73,123.2)[c]{*}
\Arc(70,210.5)(95,250,268)
\Arc(70,210.5)(95,272,290)
\Text(67,113.5)[c]{*}
\Text(73,113.5)[c]{*}
\PhotonArc(70,154.5)(48,227,313){-1}{6.5}
\Arc(70,60.5)(39,41,85)
\Arc(70,60.5)(39,95,139)
\Text(67,97.2)[c]{*}
\Text(73,97.2)[c]{*}
\Arc(70,-6)(95,70,88)
\Arc(70,-6)(95,92,110)
\Text(67,86.8)[c]{*}
\Text(73,86.8)[c]{*}
\Arc(70,174.5)(95,250,268)
\Arc(70,174.5)(95,272,290)
\Text(67,77.5)[c]{*}
\Text(73,77.5)[c]{*}
\PhotonArc(70,118.5)(48,227,313){-1}{6.5}
\end{axopicture}\vspace{3mm}
\caption{A melon-like rendition of fig.~\ref{melon}.}
\label{melonize}
\end{figure}

While the analysis of the dominant diagrams we have presented is very recognizable in its structure for anyone familiar with melonic dominance in random tensor theory \cite{melon1,melon2,melon3}, the hourglass-like graphic representation of the diagrams that we have relied upon does not superficially look like the traditional melonic diagrams. This can be remedied by bending the straight lines and positioning the stars belonging to the same chain (within the dominant contribution) next to each other. This results in a recognizable melonic shape of the graph, as in fig.~\ref{melonize}. We note, furthermore, that one should not expect a direct comparison of the coefficient in (\ref{lambdamax}) with random matrices ($q=1$), since expectation values for Gaussian ensembles of random matrices are dominated by planar diagrams, which form a bigger class than melonic diagrams relevant for Gaussian random tensors.

\subsection{$1/N$ suppression of the variance}

In the above, we have focused on the expectation value (\ref{Gauss}) and used it to supply an estimate (\ref{lambdamax}) for the largest eigenvalue of a Gaussian random tensor. An interesting question is whether this largest eigenvalue becomes determined with certainty when $N$ is large. This is the case for random matrices, where the largest eigenvalue sits almost surely at the right edge of the Wigner semi-circle. It is rather natural to expect a similar behavior for tensors.

While we do not have techniques at hand that are nearly as refined as what is available for random matrices, we can still estimate the variance of $x^{(p)}_kx^{(p)}_k$ by computing the expectation value $\langle x^{(p)}_kx^{(p)}_kx^{(p)}_lx^{(p)}_l\rangle_C$. Before the $C$-averaging, the corresponding diagram simply consists of two disconnected copies of the $(p,p)$-hourglass diagram. Averaging over $C$ is implemented by contracting the black vertices as before. However, as explained under (\ref{KKtilde}), to obtain the maximal power of $N$, the wavy lines pairing the black vertices should never reconnect any disconnected components of the graph. This means that, at leading order in $N$, the result is simply a product of the $C$-averages of each of the two hourglass diagrams, which is the same as
\beq
\langle x^{(p)}_kx^{(p)}_kx^{(p)}_lx^{(p)}_l\rangle_{\rule{0mm}{3.2mm}C}=\Big(\langle x^{(p)}_kx^{(p)}_k\rangle_{\rule{0mm}{3.2mm}C}\Big)^2.
\eeq
In other words, the variance of $x^{(p)}_kx^{(p)}_k$ is always subleading at large $N$. This strongly suggests that $\lambda_{max}$ becomes narrowly concentrated at large $N$, since the growth of $x^{(p)}_kx^{(p)}_k$ is very sensitive to $\lambda_{max}$, and any variance in $\lambda_{\max}$ would have induced a large variance of $x^{(p)}_kx^{(p)}_k$. Attempt to prove this statement rigorously would, however, require techniques outside the scope of our present treatment. (This concentration phenomenon may be seen as an example of large $N$ factorization, familiar from quantum field theory \cite{makeenko}.)


\section{Discussion}

We have examined the expectation value (\ref{Gauss}) representing the growth of the norm of a vector under application of the map (\ref{Cmap}) defined by a random real fully symmetric tensor with independent Gaussian components. Resorting to Feynman diagram techniques to analyze the Gaussian averages, we have identified the dominant diagrams in the limit of a large number of dimensions, exemplified by fig.~\ref{melon}.
Evaluation of the dominant diagrams produced an estimate (\ref{lambdamax}) for the largest tensor eigenvalue. In particular, the largest eigenvalue grows as the square root of the number of dimensions, as it does for random matrices.\vspace{1mm}

\noindent Two evident mathematical questions arise in relation to our derivations:\vspace{-3mm}
\begin{enumerate}[leftmargin=*]
\item Can one provide a rigorous justification for extracting the estimate (\ref{lambdamax}) from the growth of the norm (\ref{Gauss})?\vspace{-3mm}
\item Can one move away from Gaussian ensembles without upsetting our estimate (\ref{lambdamax})? Indeed, powerful universality results are known for eigenvalue properties of random matrices \cite{TV}. 
\end{enumerate}\vspace{-3mm}
The first question could benefit from complementary investigations of random tensor eigenvalues from different perspectives, in particular, following the approaches based on the Kac-Rice formula \cite{ABC}.
This could provide rigorous understanding in what precise sense eigenvalues greater than (\ref{lambdamax}) are uncommon. Note that approaches based on the Kac-Rice formula only pin down the numerical magnitude of the eigenvalues without saying anything about how difficult it is to find each particular eigenvalue through systematic scanning (the size of the basin of attraction of an eigenvalue for the power iterations we have employed is one way to quantify accessibility of different eigenvalues). In relation to the second question, powerful results on universality are known for random tensors \cite{univ} and suggest that our findings are robust in a large class of random tensor ensembles with independently distributed entries. (As an aside, the maximal H-eigenvalue of random tensors with entries taking values $\pm 1$ treated in \cite{cooper} displays a dependence on $N$ different from our results on E-eigenvalues.)

As to further applications of diagrammatic techniques and melonic dominance, one could effectively compute averages of quantities polynomial in the random tensor. In particular, polynomial expressions for generalizations of traces and for resolvents have been developed in \cite{MSh}. This may open a way to compute averages of characteristic polynomials of tensors, in analogy to the corresponding work on random matrices \cite{charrand}. Needless to say, much remains to be explored.\vspace{5mm}

\noindent {\bf Note added:} A few weeks after this article was released as a preprint, independent closely related work \cite{wiggen} appeared. This work introduces a tensor-based distribution that reduces to the standard Wigner distribution for random matrix eigenvalues when the tensors are of rank 2. Just like in the above presentation, the construction is motivated by exploring eigenvalues of random tensors. It would be interesting to understand the relation between the two approaches in more detail.


\section*{Acknowledgments}

I have benefitted from discussions with Charles Bordenave, Beno\^{i}t Collins, Luca Lionni, Ion Nechita, Gillaume Valette and especially Vincent Rivasseau. Luca Lionni has  also contributed valuable comments on the manuscript.
This research is supported by CUniverse research promotion project at Chulalongkorn University (grant CUAASC).



\end{document}